\documentclass[preprinty]{aastex}
\usepackage{longtable}
\usepackage{graphicx,amssymb,mathrsfs,amsmath}
\usepackage{lscape}
\usepackage{color}

\begin{document}

\title{A SOLAR ERUPTION DRIVEN BY RAPID SUNSPOT ROTATION}

\author{GUIPING RUAN\altaffilmark{1,2,3},
        YAO CHEN\altaffilmark{1},
        SHUO WANG\altaffilmark{3},
        HONGQI ZHANG\altaffilmark{2},
        GANG LI\altaffilmark{4},
        JU JING\altaffilmark{3},
        JIANGTAO SU\altaffilmark{2},
        XING LI\altaffilmark{5},
        HAIQING XU\altaffilmark{2},
        GUOHUI DU\altaffilmark{1},
        and HAIMIN WANG\altaffilmark{3}
}

\affil{1 Institute of Space Sciences and School of Space Science and
  Physics, Shandong University, Weihai 264209, China;
  yaochen@sdu.edu.cn} \affil{2 Key Laboratory of Solar Activity,
  National Astronomical Observatories, Chinese Academy of Sciences,
  Beijing 100012, China} \affil{3 Space Weather Research Laboratory, Center
  for Solar-Terrestrial Research, NJIT, Newark, NJ07102, USA}
  \affil{4 Department of Physics and CSPAR,
  University of Alabama in Huntsville, Huntsville, AL 35899, USA}
\affil{5 Institute of Mathematics and Physics, University of
  Aberystwyth, UK}

\begin{abstract}
We present the observation of a major solar eruption that is
associated with fast sunspot rotation. The event includes a
sigmoidal filament eruption, a coronal mass ejection, and a GOES
X2.1 flare from NOAA active region 11283. The filament and some
overlying arcades were partially rooted in a sunspot. The sunspot
rotated at $\sim$10$^\circ$ per hour rate during a period of 6
hours prior to the eruption. In this period, the filament was
found to rise gradually along with the sunspot rotation. Based on
the HMI observation, for an area along the polarity inversion line
underneath the filament, we found gradual pre-eruption decreases
of both the mean strength of the photospheric horizontal field
($B_h$) and the mean inclination angle between the vector magnetic
field and the local radial (or vertical) direction. These
observations are consistent with the pre-eruption gradual rising
of the filament-associated magnetic structure. In addition,
according to the Non-Linear Force-Free-Field
reconstruction of the coronal magnetic field, a pre-eruption
magnetic flux rope structure is found to be in alignment with the
filament, and a considerable amount of magnetic energy was
transported to the corona during the period of sunspot rotation.
Our study provides evidences that in this event sunspot rotation
plays an important role in twisting, energizing, and destabilizing
the coronal filament-flux rope system, and led to the eruption. We
also propose that the pre-event evolution of $B_h$ may be used to
discern the driving mechanism of eruptions.
\end{abstract}


\section{Introduction}
Solar eruptions, including solar flares, coronal mass ejections
(CMEs), and filament eruptions, are spectacular energy release
phenomena that occur in the solar atmosphere. They often lead to
catastrophic impacts on the near-Earth space environment. They are
generally believed to be a result of the rapid release of magnetic
energy stored in highly-stressed/twisted magnetic structures of
the corona (e.g., Forbes 2000; Low 2001). The magnetic energy is
transported into the corona via slow photospheric footpoint
motions, e.g., emergence, shearing, twist, etc., in a relatively
long period of time, comparing to the time scale of an eruption.
Among various forms of photospheric motions, sunspot rotation,
first observed a century ago by Evershed (1910), has been
considered to be an important process and has been studied extensively
(e.g., Stenflo 1969; Barnes \& Sturrock 1972; Ding et al.
1977, 1981; Amari et al.  1996; Tokman \& Bellan 2002;
T$\ddot{o}$r$\ddot{o}$k \& Kliem 2003; Brown et al. 2003; Regnier
et al. 2006; Yan et al. 2007, 2008a, 2008b, 2009; Su et al. 2010).

Previous studies confirmed the important role played by sunspot
rotation in transporting energy and helicity from below the
photosphere into the corona with quantitative calculations (e.g.,
Kazachenko et al. 2009; Vemareddy et al. 2012), and revealed some
temporal and spatial association of sunspot rotation with solar
flares on the basis of observational data analysis (e.g., Zhang,
Li \& Song 2007; Zhang, Liu \& Zhang 2008; Yan et al. 2007, 2008a,
2008b, 2009; Jiang et al. 2012). There also exist a number of
magnetohydrodynamic (MHD) studies examining the consequence of
twisting a flux rope structure which is confined by overlying
magnetic arcades (e.g., Amari et al. 1996; T$\ddot{o}$r$\ddot{o}$k
et al. 2003). In a latest study, T$\ddot{o}$r$\ddot{o}$k et al.
(2013) examined the role of twisting the overlying arcades in the
onset of a CME using a flux rope model. These studies showed that
the CME can be triggered by twisting either the core flux rope
structure or the overlying coronal fields, thus established the
importance of sunspot rotation in the eruption process from a
theoretical perspective. On the other hand, observational studies
connecting sunspot rotation with CMEs remain elusive.

Here we present a case study of the evolution of a sigmoidal
filament which has roots in a rotating sunspot. The study,
involving multi-wavelength imaging and vector magnetic field data
from the Solar Dynamics Observatory (SDO), provides a rare case
revealing the role of sunspot rotation being as not only a general
energy transport process but also a direct driving process that
leads to the eventual flare, CME, and filament eruption.

\section{Observation}
We analyzed the multi-wavelength imaging data provided by the
Atmospheric Imaging Assembly (AIA; Lemen et al. 2012) and the
vector magnetic field and continuum intensity data by the
Helioseismic and Magnetic Imager (HMI; Schou et al. 2012) on board
the SDO spacecraft for the NOAA active region (AR) 11283 between
2011 September 3 and September 8. The AR was located N14W15 at
16:00 UT on September 6, close to the disk center. AIA observes
the Sun in 10 different wavebands, covering a wide range of
temperatures and reveals physical processes at various layers of
the solar atmosphere. The data are taken with a pixel size of
0.6$^{\prime}$$^{\prime}$ and 12s cadence. For our study, we only
analyze the AIA observations at the 304\,\AA{} (HeII, T$\sim$ 0.05
MK) to follow the dynamics of the cool filament and the 94\,\AA{}
(FeXVIII, T$\sim$6.3 MK) observation to trace the hot eruptive
structures. The processed disambiguated HMI vector magnetic field
data are of 12-minute cadence at a 0.5$^{\prime}$$^{\prime}$ pixel
resolution, provided by the HMI team (see
ftp://pail.stanford.edu/pub/HMIvector2/movie/ar1283.mov for the
corresponding movie). These vector magnetogram data have been
de-rotated to the disk center, and remapped using a Lambert equal
area projection (Calabretta \& Greisen 2002; Thompson 2006). The
field vectors are then transformed to Heliographic coordinates
with projection effect removed (Gary \& Hagyard 1990, also see Sun
et al., 2012).

In Figure 1, we present the intensity map (panel a) and HMI vector
magnetogram (panel b) at $\sim$22:00 UT on September 6,
just before the X2.1 flare.
We show the local vertical (i.e., radial) magnetic
field component ($B_z$) in white and black for positive and
negative polarities. The color-coded arrows in panel (b) represent
the horizontal magnetic field $B_h$, which is the component
parallel to the solar surface (i.e., $B_h = \sqrt{(B_x^2 +
B_y^2)}$, where $x$ and $y$ represent  two orthogonal directions
in the plane of the solar surface). The yellow curve represents the
magnetic polarity inversion line (PIL). From the temporal
evolution of the HMI vector magnetic field, this AR is
characterized by an emerging positive polarity sunspot. The
emergence started from the heliographic location N13E28 near the end
of September 3 and was the dominant process in the first two days.
After that, the AR developed into a $\beta\gamma\delta$ magnetic
complexity.  Since early September 6, the emerged sunspot
exhibited an apparent clockwise rotation, as well as a slow
westward shearing motion along the PIL. The rotation direction is
consistent with the right-handed twist of the horizontal field as
can be seen in panel b. Near the PIL, this field component is
almost parallel to the PIL indicating the presence of strong
magnetic shear.

Many flares have been produced by this AR from September 3 to 7.
Among them, three big flares were observed on September 6 and 7
with GOES SXR flare classes being M5.3, X2.1, and X1.8. Their
peaking times were 1:50 UT and 22:20 UT on September 6, and 22:38
UT on September 7, respectively. The sunspot rotation can be
discerned a few hours before and after the M5.3 flare. It then became
harder to trace until at $\sim$16:00 UT, 6 hours before the X2.1 flare,
when two magnetic tongues (c.f., Schmieder et al. 2007) formed, providing
excellent tracer to the rotation. We focus our analysis in this 6-hour
period to examine the role of sunspot rotation in the onset of the
eruption associated with the X2.1 flare.

It is important to understand the topology of the coronal magnetic
structure and how important the rotation in the coronal energy
accumulation process. To achieve this, we reconstructed the
three-dimensional (3D) coronal magnetic field using the nonlinear
force-free field (NLFFF) extrapolation method developed by
Wiegelmann (2004) and (Wiegelmann et al. 2006) on the basis of HMI
data. Details of the method are presented in the Appendix.

\section{Results and interpretation}
The X2.1 flare started at 22:12 UT, peaked at 22:20 UT, and ended at
22:50 UT according to the GOES x-ray (1-8\,\AA{}) light curve shown in
Figure 2. The pre-flare (22:06 UT) and flare-peaking (22:20 UT) images
observed in the 94\,\AA{} bandpass have been shown in panels c and d
of Figure 1 and the accompanying animation. The pre-flare hot
structures exhibited an arcade connecting the northern and southern
ends of the eruptive structure, two sets of arcade loops of different
size, and highly twisted structures at the north-west part of the
image. The large bright area in the post-flare 94\,\AA{} image
indicates a strong heating there.

The flare was accompanied by a halo CME travelling at a linear
speed of 575 km $s^{-1}$ according to the CDAW (Coordinated Data
Analysis Workshops) catalog of the LASCO data (Brueckner et al.
1995). The eruption was also observed with the STEREO spacecraft
(Howard et al. 2008) as a limb event. Panels e and f of Figure 1
present two subsequent images at $\sim$22:26 UT and 22:31 UT
observed by COR1 and EUVI aboard STEREO-B. We see that the CME
front moved $\sim$0.5 R$\odot$ within 5 minutes yielding a speed
of $\sim$1200 km s$^{-1}$, much faster than that measured with
LASCO. This is mainly due to the projection effect and the CME
deceleration during its propagation to the outer corona.

Figure 3 presents sequences of the sunspot (a-c) and the filament
(d-f) morphological evolutions. The contours in panels b and e
represent the $\pm$350~G level of $B_z$ at $\sim$19:00 UT. We can
see from this figure and the online animation that the dominant
motion in this period was the sunspot rotation. The sunspot
developed co-rotating magnetic tongues at $\sim$16:00 UT on
September 6. This allows a quantitative determination of the
rotation rate. To do this, we present in Figure 3g the $r-\theta$
time-slice plot of the sunspot for the time range of 10:00 UT to
24:00 UT. The $r-\theta$ plot was produced by retrieving images
along two semi-circular slices with a radius of
$\sim$3.5$^{\prime}$$^{\prime}$ and 3$^{\prime}$$^{\prime}$
around the sunspot center (which is in motion), and
stack them over time. The $\theta=0^\circ$ is along the northward
(upward) direction. The angle increases in the clockwise
direction. We use two slices so that we can examine both tongues
simultaneously.

The $r-\theta$ plot reveals features consistent with the above
description of the sunspot rotation. We can see that the sunspot
rotated by $\sim$~60$^\circ$ in the 6 hours prior to the flare,
with an average rotation rate $\sim$10$^\circ$ per hour. After the
flare (peaking time shown by the blue vertical line), the sunspot
experienced a sudden morphological change and the rotation became
hard to track. In comparison with events reported earlier (e.g.,
Zhang et al. 2007; Yan et al. 2012), our event can be regarded as
a fast rotation one. Along with the rotation, the sunspot center
moved westward by $\sim$2$^{\prime}$$^{\prime}$ in the 6 hours.
Comparing to the fast rotation, the shearing motion seems
to be insignificant, which is therefore presumed to play a less
important role in the onset of the eruption.

From panels d-f of Figure 3, the entire filament structure
exhibited a highly curved pattern with two segments. The southern
and the northern segments were disconnected from each other at
both ends. The southern filament erupted first which was followed
by the eruption of the northern filament. In this study, we focus
only on the southern filament. It presented a highly-curved
sigmoidal morphology. Since its northern end was mostly rooted in
the sunspot penumbra region, we suggest that the sunspot rotation
was directly related to the dynamics of this filament.

There were clear filament morphological changes during the 6-hour
period. According to Figure 3 and the accompanying online
animation, the filament exhibited separated filamentary structures
which seem to be twisted around each other. The filament became
more bulging after 20:00 UT. At $\sim$22:00 UT, the filament
already started to rise rapidly before the start of the flare
(22:12 UT). The filament motion can be viewed from the time-slice
plot shown in panel h. The slice is drawn in panel d as a white
line. The white-dotted line in panel h is drawn to indicate the
moving filament. As can be seen, the motion of the filament along the
slice was hardly measurable before 16:00 UT, and was clear
from 16:00 UT to 21:00 UT, during which it moved a distance
of $\sim$5 Mm ($\sim$7$^{\prime}$$^{\prime}$).

It has to be noted that the above measured moving distance
consists of contributions from both the radial (or vertical) and
horizontal motions of the filament, and it is generally difficult
to disentangle them due to the projection effect. Nevertheless, we
can estimate the maximum rising distance of the filament by
assuming a pure radial motion. With this assumption, the
deprojected motion ($\vartriangle$$R$) of the filament can be
given by $\vartriangle$$R = \vartriangle$$r {R_\odot \over r}$,
where $\vartriangle$$r$ and $r$ are the filament moving distance
and the distance from the filament center to the solar center as
measured in the projection plane. The real distance from the
filament to the solar center is approximated by the solar radius
assuming that the initial filament height is negligible comparing
to the solar radius. A schematic showing the relationship between
these parameters is shown in Figure 4. According to the AIA data,
we have $r \sim 0.34 R_\odot$ at 16:00 UT and $\vartriangle$$r
\sim 7^{\prime\prime}$. This leads to a maximum rising height of
$\vartriangle$$R$$\sim 20^{\prime\prime}$.

In the pre-event process, several temporary and persistent
brightening structures were observed in the 94\,\AA{} bandpass
(see the animation accompanying Figure 1), indicating the
existence of reconnections. These reconnections can release part
of the accumulated energy and affect the dynamics and morphology
of the filament.

To further explore the details of magnetic field evolution, we
display the distributions of $B_h$ at 22:00\,UT (panel a) and
22:36\,UT (panel b) in Figure 5. It can be seen that $B_h$
increased rapidly after the flare, in agreement with previous
studies (e.g., Wang et al. 1994, 2010; Liu et al. 2012; Wang et
al. 2012; Sun et al. 2012). The post-flare $B_h$ contour
observed at 22:36 UT at a level of 1250 G is plotted in both
panels, outlining the major region of the flare-induced $B_h$
enhancement, which is referred to as region R hereafter. It can be
seen that region R is underneath the filament and across the PIL.
In this study, we focus on the variation of the pre-flare
photospheric field in the above region defined by the post-flare
$B_h$ enhancement. Note that magnetic field measurement during
flare time (within $\sim$30 minutes in general) is less accurate
than before and after the flare (e.g., Qiu \& Gary, 2003).

The temporal evolutions of the average $B_h$ and the positive and
negative $B_z$ in R are plotted in Figure 6, as the black-solid,
blue-dotted, and dashed lines, respectively. The error bars of the
$B_h$ and $B_z$ data shown in this figure are given by $3\sigma$
where $\sigma$ is the standard deviation of the HMI data obtained
from a nearby quiet-sun region. Also plotted are the inclination
angle of the vector photospheric magnetic field ($\theta_B$)
(i.e., the angle between the local vertical direction and the
vector magnetic field) in green and the total flux in red-dotted.

It can be seen that before the sudden changes of $B_h$ and
$\theta_B$, there were gradual but steady decreases of both
quantities. This trend was especially clear during the 6-hour
period between 16:00 UT and 22:00 UT. Indeed, the average $B_h$
decreased consistently by 15\% from about 1185 G at 16:00 UT to
1009 G at 22:12 UT. In comparison, both the absolute value and
variation of $B_z$ were much smaller than that of $B_h$ during the
6 hours before the flare. The positive $B_z$ increased from 350 G
to 410 G, while the negative one changed from $-$160 G to $-$153 G
during the same period. The total flux (the red dotted line)
presented a slow yet steady increase with no apparent change of
increasing rate during the period of sunspot rotation (i.e., after
16:00 UT). On the other hand, the average $\theta_B$ changed
persistently from $\sim$70$^\circ$ to $\sim$63$^\circ$ during the
6 hours of sunspot rotation. This suggests that the relevant
magnetic structures became more vertical. The total magnetic field
strength (not shown here) decreases gradually in a manner similar
to that of $B_h$ since $B_h$ is much stronger than $B_z$. It is
expected that when a magnetic structure rises into the corona it
will expand into a larger volume. This will result in a decrease
of both the total and the horizontal magnetic field strength,
consistent with our observation.

Selected field lines of NLFFF reconstructions are shown in Figures
5c and 5d. The location of the low-lying twisted magnetic
structure, i.e., a flux rope structure, co-aligned with the
southern filament. Note that the magnetic topology of this event
was also reconstructed and studied by Jiang \& Feng (2013), Jiang
et al. (2013a, 2013b), and Feng et al. (2013). They have presented
similar reconstruction results as shown here. We can see that
these field lines and some overlying magnetic arcades were rooted
in the rotating sunspot, agreeing with the observation shown in
Figure 3. Thus, the sunspot rotation may affect both the twisted
filament-flux rope structure and the overlying arcade. After the
flare, the field lines in the filament location became shorter and
less twisted indicating a relaxed energy state.

We plot in Figure 7(a) the temporal profile of the total
energy of the reconstructed magnetic field in a sub-volume with a
bottom shown as the blue square in Figure 7(b) and the same height
as that used for the NLFFF reconstruction. The sub-domain is
selected to focus on the smaller region of eruption. An estimate
of the total magnetic energy in the whole reconstruction domain
yields a very similar profile. We see that the total magnetic
energy in this sub-domain shows a rapid increase after 16:00 UT,
which is the starting time of the apparent sunspot rotation, and
an abrupt decline during the flare. The energy increase from 16:00
UT to $\sim$22:00 UT is about 3 $\times10^{31}$ erg, which is
capable of energizing a major solar event (e.g., Vourlidas et al.,
2002) and therefore probably important to the onset of the
following eruption. This indicates that the sunspot rotation,
which is a major dynamical feature of the active region, is
important to the pre-eruption energy storage in the corona. A
detailed study on the energetics of this event, including the
estimates of the free magnetic energy before and after the
eruption, the thermal and nonthermal energies for the flare, the
CME kinetic and potential energies, as well as the partition of
the released magnetic energy between the CME and the flare, has
been presented by Feng et al. (2013). They show that the flare and
the CME may have consumed a similar amount of magnetic free energy
within the estimate uncertainty.

One likely driving mechanism of the eruption in our event is
illustrated in the schematics shown in Figure 8. The white
structure with two extending tongues indicates the rotating
sunspot with the rotating direction denoted by the arrows. The
long twisted field lines along the PIL represent the magnetic
structure associated with the filament, representing the
filament-carrying flux rope structure whose chirality is
consistent with the direction of sunspot rotation. The flux rope
and a part of the overlying arcades are rooted in the rotating
sunspot. Thus, the sunspot rotation can  directly transport energy
and helicity into the coronal flux rope system.
T$\ddot{o}$r$\ddot{o}$k et al. (2013) proposed a novel mechanism
for CME eruption induced by the expansion of the overlying arcades
that are rooted in a rotating sunspot. Our study provides
observational evidences supporting their scenario. Note that our
observations show that both the flux rope-filament structure and
the overlying arcades were twisted by the rotation of the sunspot.
Both twistings may play a role in driving the eruption in our
event, and it is not possible to disentangle them. From Figure 8a
to 8c, the sunspot rotates about $\sim$60$^\circ$, as indicated by
the locations of the two tongues. Correspondingly, the central
part of the twisted field lines expands and moves higher and the
overlying arcades become more vertical. These features agree with
the observations of the filament rising and the gradual decreases
of both the horizontal component and the inclination angle of the
photospheric magnetic field. In short, the observational features
in our event can be understood with a flux rope CME driven by a
persistent sunspot rotation, as schematically illustrated here.

\section{Summary}
We present observations of a sunspot rotation before a major solar
event consisting of a fast CME, an X2.1 flare, and a filament
eruption. We suggest that this pre-eruption rotation is not only
transporting energy to the corona, but also playing a dynamic and
critical role in leading to the eruption. Our suggestion is based
on the data analysis results, which are summarized below. Firstly,
the sunspot rotation was the dominant motion in the 6 hours before
the flare. The rotation rate was $\sim$10$^\circ$ per hour,
considerably faster than some previous observations (e.g., Zhang,
Liu, \& Zhang 2008). Secondly, the filament and part of the
overlying arcades were rooted in the rotating sunspot, and the
filament exhibited an apparent gradual rising motion along with
the sunspot rotation. This provides a possible dynamical link
between the sunspot rotation and the filament dynamics as well as
the resultant eruption. Thirdly, the evolutions of both the
photospheric horizontal field and the magnetic field inclination
angle agree with the gradual rising of the magnetic structure that
supports the filament. Last, using the NLFFF method of coronal
magnetic field reconstructions, we find the presence of a
well-developed twisted flux rope structure associated with the
filament and a considerable amount of magnetic energy increase
during the sunspot rotation period. These results highlight the
importance of sunspot rotation to the energy storage and the onset
of the eruption.

The evolution of photospheric magnetic field is essential to both
the energy build-up and the triggering of a solar eruption.
 Many studies have focused on rapid changes of $B_h$ induced by the
flare (e.g., Wang et al. 1994, 2010; Liu et al. 2012; Wang et al.
2012). However, the detailed pre-flare evolution of this field
component has not received sufficient attention. Such evolution
would carry important information of the energy storage and
eruption onset process. In our study, we found that $B_h$ in the
area underneath the filament decreased gradually in hours before
the flare. This is related to the gradual ascending of the
filament-flux rope structure. Our analysis indicates that this is
associated with the rapid sunspot rotation. On the contrary,
studies of another active region (AR11158) revealed that $B_h$
there increased gradually in a similar time period prior to the
flare (Liu et al. 2012; Wang et al. 2012). Those studies deduced
that the corresponding eruptions were driven by tether-cutting
reconnection (Moore et al. 2001) of two approaching magnetic
loops. The pre-flare footpoint separation of these loops can
explain the gradual increase of $B_h$. By comparing the different
behavior of $B_h$ and corresponding understanding of the eruption
mechanism, we suggest that the pre-flare variation of $B_h$ can be
taken as a clue to discern the eruption mechanism: A gradual
decrease of $B_h$ may be a precursor for an eruption in terms of
the flux rope instability, while an increase of $B_h$ may be the
precursor for tether-cutting reconnection. This needs further
clarifications in future studies.

\acknowledgements We thank SDO/HMI and SDO/AIA science teams for
the free access to the data. We are grateful to Dr. Guangli Huang,
Jie Zhang, Yang Liu, and Xudong Sun for useful discussions
and the anomalous referee for constructive comments. GPR
thanks the NJIT people for their hospitality during her visit.
This work was supported by the 973 program NSBRSF 2012CB825601,
NNSFC grants 41331068, 41274175 and U1331104 to SDUWH, by NSF
grants AGS 1153226, 1153424 and 1250374 to NJIT, AGS1135432 and
ATM0847719 to UAH, and by NNSFC grants (11373040, 11003025,
41174153, 11178005, 11221063) and the Specialized Research Fund
for State Key Laboratories to NAOC. GPR was also supported by the
CAS Key Laboratory of Solar Activity.

\appendix

\section{NLFFF extrapolation method of the coronal magnetic field}

The coronal magnetic field was reconstructed using the
extrapolation method developed by Wiegelmann (2004). The code of
the method was provided by Thomas Wiegelmann. The HMI magnetograms
were preprocessed to remove most of the net Lorentz force and
torque from the data so as to be more consistent with the
force-free assumption (Wiegelmann et al. 2006). The extrapolation
was performed using 2$\times$2 rebinned magnetograms within a box
of 217$\times$185$\times$145 Mm$^3$ at the 12 minute cadence. The
corresponding grid number is taken to be 300$\times$256$\times$200
with a uniform spacing of 1.0$^{\prime}$$^{\prime}$.

The method employs a weighted optimization approach which
minimizes a joint measure for the Lorentz force density and the
divergence of the field throughout the computational domain
(Wheatland et al. 2000), which is represented by the optimization
integral $L$. The performance of the method is further evaluated
by calculating the average dimensionless field divergence $f$ and
the current-weighted average of $\sin\theta$ (CWsin) where
$\theta$ is the angle between the vector magnetic field
$\textbf{\emph{B}}$ and the current density $\textbf{\emph{J}}$
($0^{\circ}\leq\theta\leq180^{\circ}$) (c.f., Schrijver et al.
2006, 2008; Metcalf et al. 2008; DeRosa et al. 2009). The
optimization measure $L$ is defined as
$$
L=<{\omega}_f(\boldsymbol{r})B^{-2}{\lvert{(\nabla \times
\boldsymbol{B})\times\boldsymbol{B}}\rvert}^2>
+<{\omega}_d(\boldsymbol{r}){\lvert{\nabla \cdot
    \boldsymbol{B}}\rvert}^2>
    \eqno{(A1)}
$$
where the angle bracket denotes the mean value within the domain,
the first and second parts of $L$ represent a measure of the mean
Lorentz force density ($L_f$) and the mean field divergence
($L_d$), respectively. Both $\omega_f$ and $\omega_d$ are position
dependent to reduce the effect of boundary conditions. They are
fixed to be 1.0 in the center of the computational domain and drop
to 0 monotonically with a cosine profile in a buffer boundary
region that consists of 32 grid points toward the side and top
boundaries. It is found that the optimization measure $L$
decreases from an initial value of 109.6 to a final value of 11.5,
while the field divergence measure $L_d$ decreases from $\sim$47.4
to 4.0, and the Lorentz force measure $L_f$ decreases from
$\sim$62.2 to 7.5, in units of G$^2$ arcsec$^{-2}$. These values
of the optimization measure are comparable to previously reported
values for other events (e.g., Schrijver et al., 2008; Sun et al.,
2012).

The code checks whether $L(t + dt) < L(t)$ after each time step.
If the condition is not fulfilled, the time step $dt$ is reduced
by a factor of 2 and the iteration step is repeated. After each
successful iteration step $dt$ is increased by a factor of 1.01.
This allows $dt$ to become as large as possible while satisfying
the stability condition. The iteration stops if the condition
$|{\vartriangle{L_w} \over \vartriangle{t}}|/{L_w}<10^{-4}$ is
satisfied for 100 consecutive iteration steps.

The current weighted average of $\sin\theta$ is defined as
$$
\texttt{CWsin}={{\sum_{i}\lvert{J_i}\rvert \sigma_{i}} \over
{\sum_{i}\lvert{J_i}\rvert}}, \sigma_{i}={{\lvert{J_i} \times
B_{i} \rvert} \over {\lvert{J_i}\rvert \lvert{B_i}\rvert}}=\lvert
{sin{\theta_{i}}} \rvert, \eqno{(A2)}
$$
and the pointwise average of the divergence $f$ is defined by
$$
f=<|f_i|>=<{|(\nabla\cdot\textbf{B})_i| \over
(6|\textbf{B}|_i/\triangle x)}>,\eqno{(A3)}
$$
where $i$ represents the grid point and $\Delta{x}$ is the grid
spacing (c.f., Schrijver et al. 2006, 2008; Metcalf et al.
2008; DeRosa et al. 2009). For the final reconstruction results,
we find that the mean CWsin varies in a range of 0.33 - 0.41 with
an average of 0.36, and the average field divergence $\lvert f
\rvert$ varies in between 0.00072 and 0.00091 with a mean value of
0.00082.

We acknowledge that there exist other parallel NLFFF codes that
have been broadly used or evaluated by solar physics researchers
(e.g., Schrijver et al. 2006; Metcalf et al. 2008; Schrijver et
al. 2008; DeRosa et al. 2009). Given the limitation of both the
NLFFF algorithm and the vector magnetic field measurements, the
reconstruction results should be assessed with caution.

\begin{figure*}[!htbp]
\begin{minipage}{\textwidth}
\centering
\includegraphics[width=128mm]{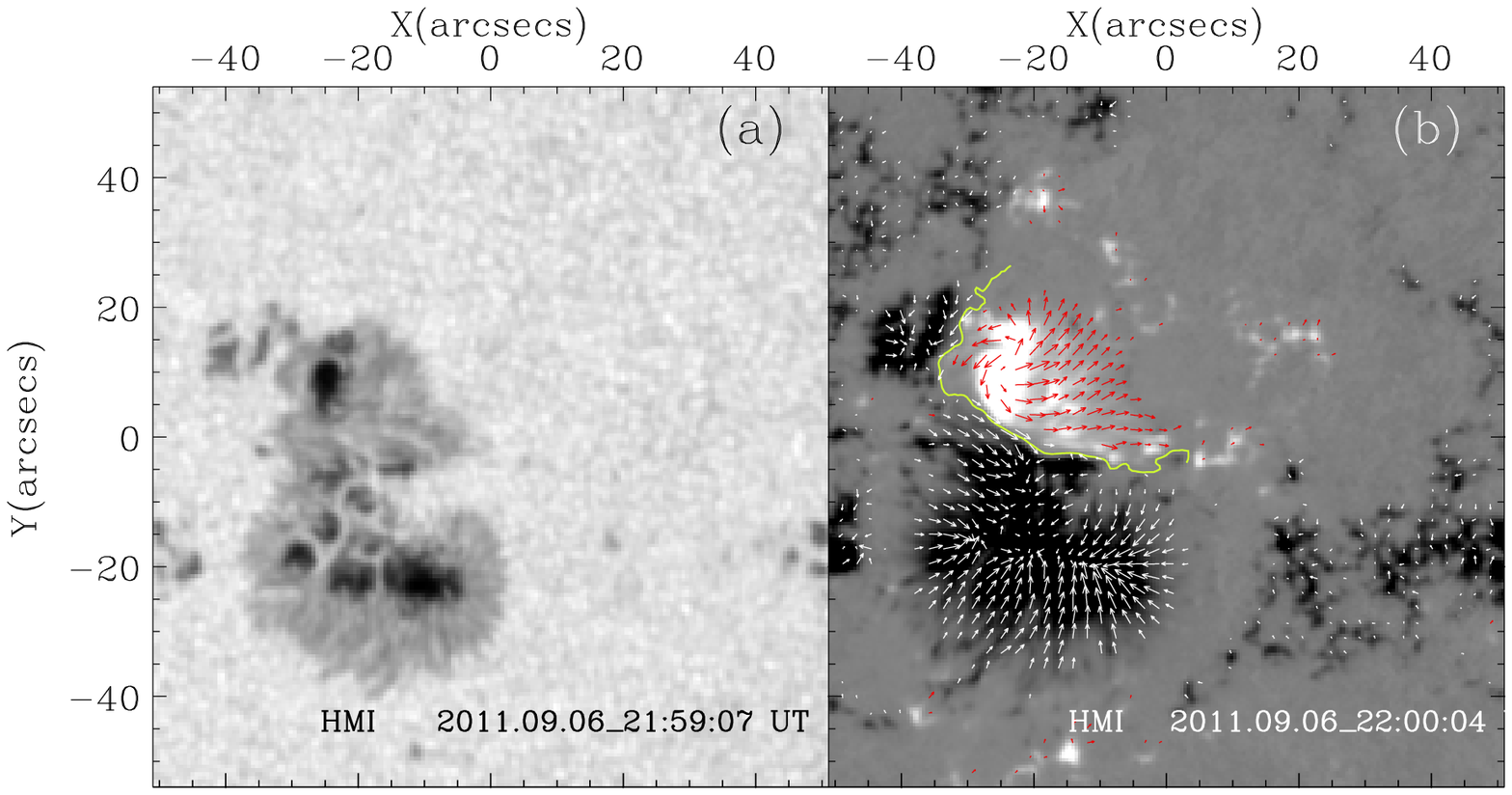}
\vspace{-2.3mm}
\end{minipage}
\begin{minipage}{\textwidth}
\centering
\includegraphics[width=128mm]{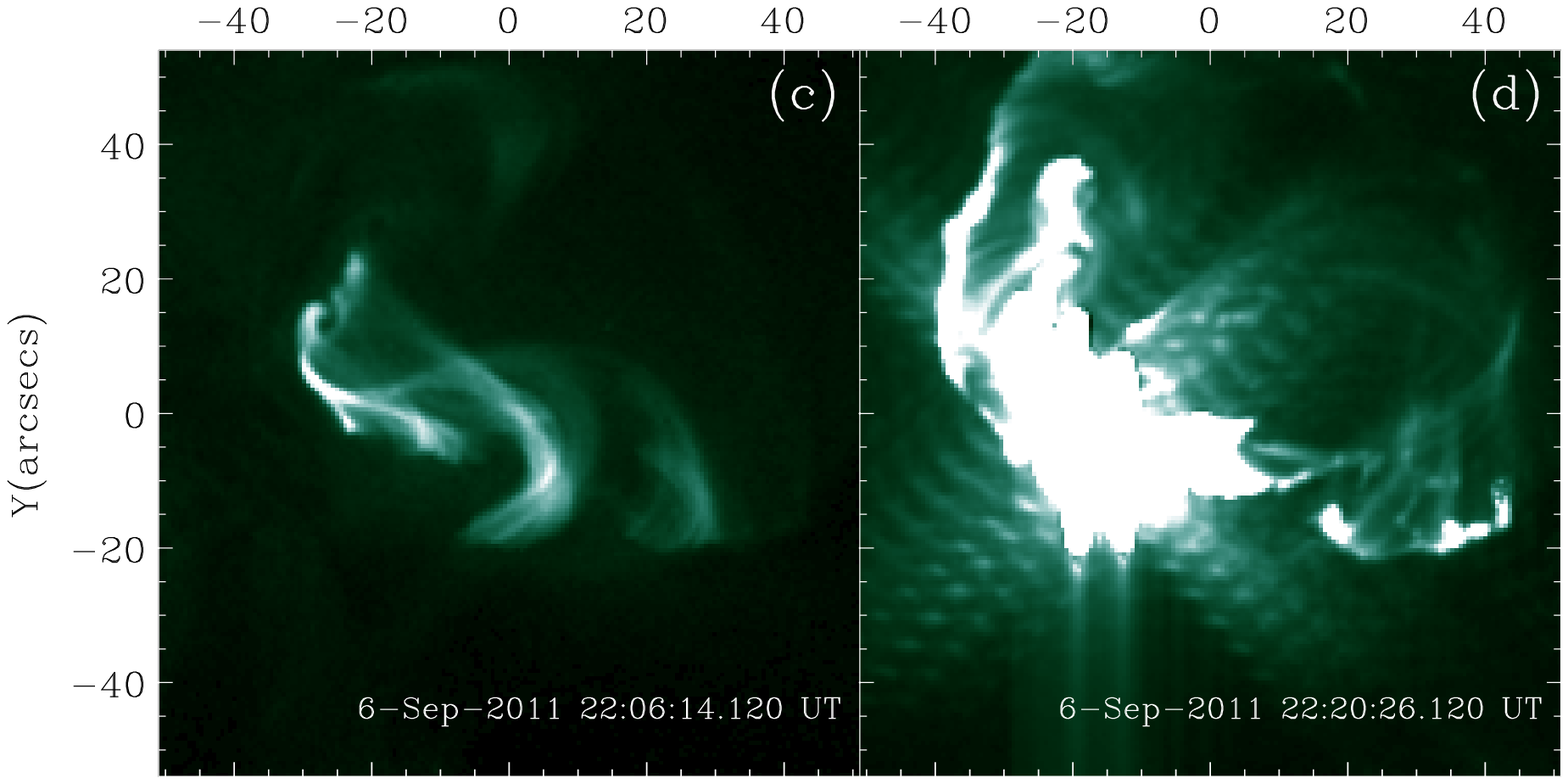}
\vspace{-0.3mm}
\end{minipage}
\hspace*{31.3mm}
\begin{minipage}{\textwidth}
\includegraphics[width=107mm]{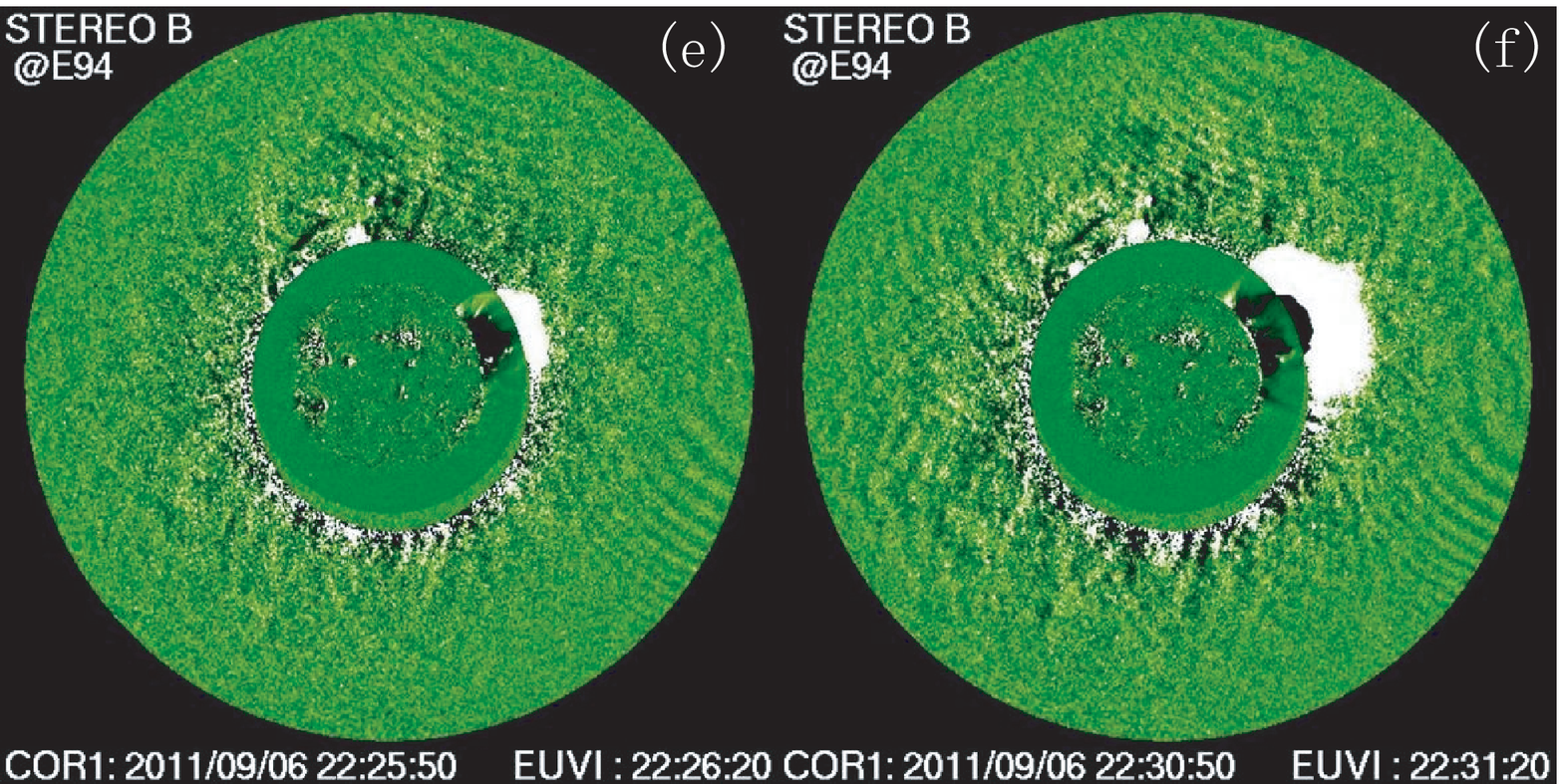}
\end{minipage}
\vspace{-7.3mm} \caption{(a, b): The HMI continuum intensity image
and vector magnetogram for the NOAA AR 11283 (N14W18) observed at
$\sim 22:00$ UT on September 6, 2011. $B_z$ is shown in white
(black) for positive (negative) polarity, $B_h$ is represented
with arrows that are color-coded according to the corresponding
$B_z$ polarities. The yellow line represents the PIL. (c, d): The
AIA 94\,\AA{} images at 22:06 UT and 22:20 UT in the same FOV as
panel a. (e, f): The CME images observed by STEREO-B. Animation of
the AIA 94\,\AA{} data and a color version of this figure are
available online.}
 \label{fig:bright}
\end{figure*}

\begin{figure*}[!htbp]
\centering
\includegraphics[width=150mm]{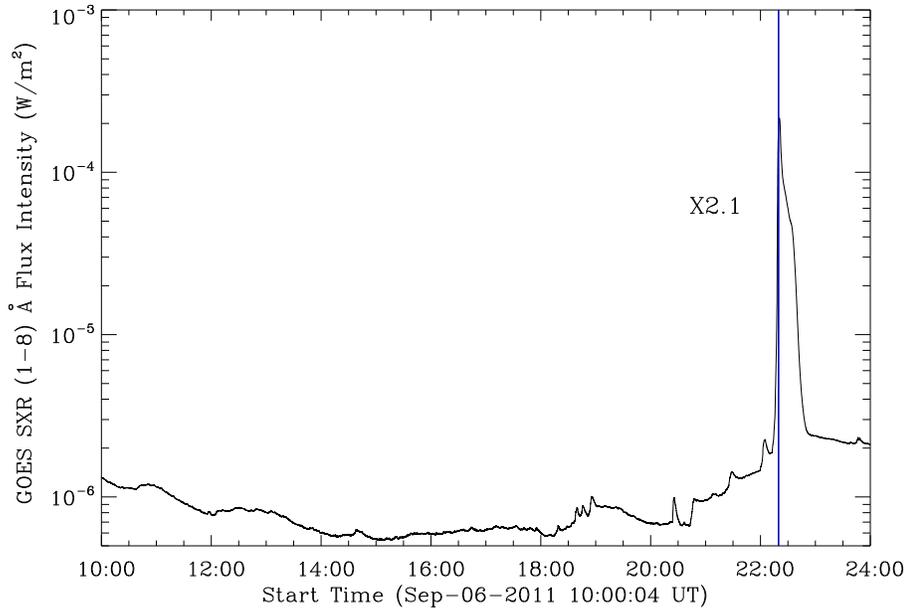}
\caption{The 1-8 \,\AA{} GOES SXR flux intensity profiles. The
blue vertical line represents the flare peaking time (22:20 UT). A
color version of this figure are available online.}
\label{fig:bright}
\end{figure*}

\begin{figure*}[!htbp]
\begin{minipage}{\textwidth}
\centering
\includegraphics[width=140mm]{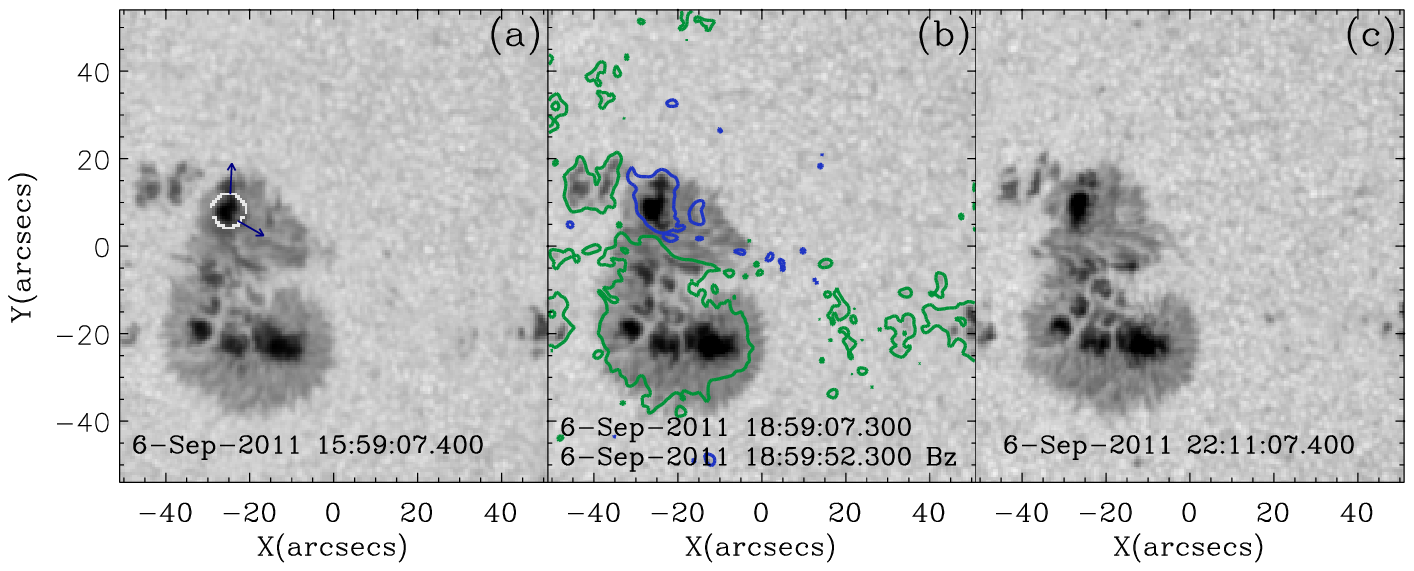}
\vspace{-2.3mm}
\end{minipage}
\begin{minipage}{\textwidth}
\centering
\includegraphics[width=140mm]{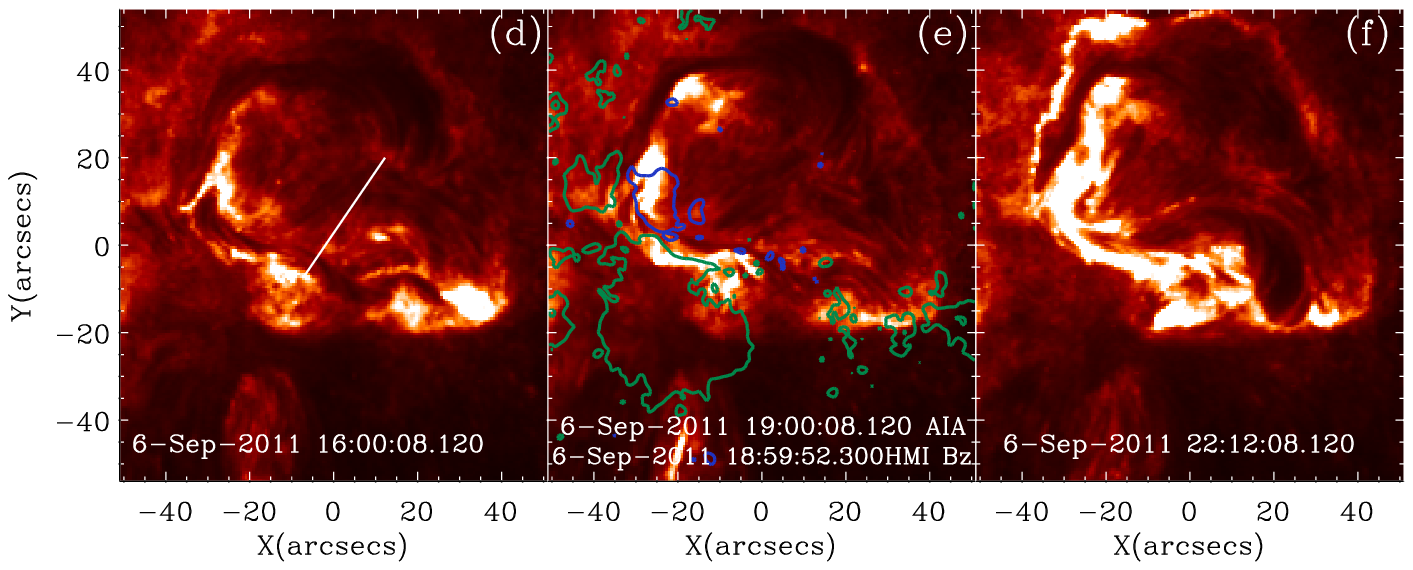}
\vspace{-0.3mm}
\end{minipage}
\hspace*{12.3mm}
\begin{minipage}{\textwidth}
\includegraphics[width=0.42\textwidth,angle=0,clip]{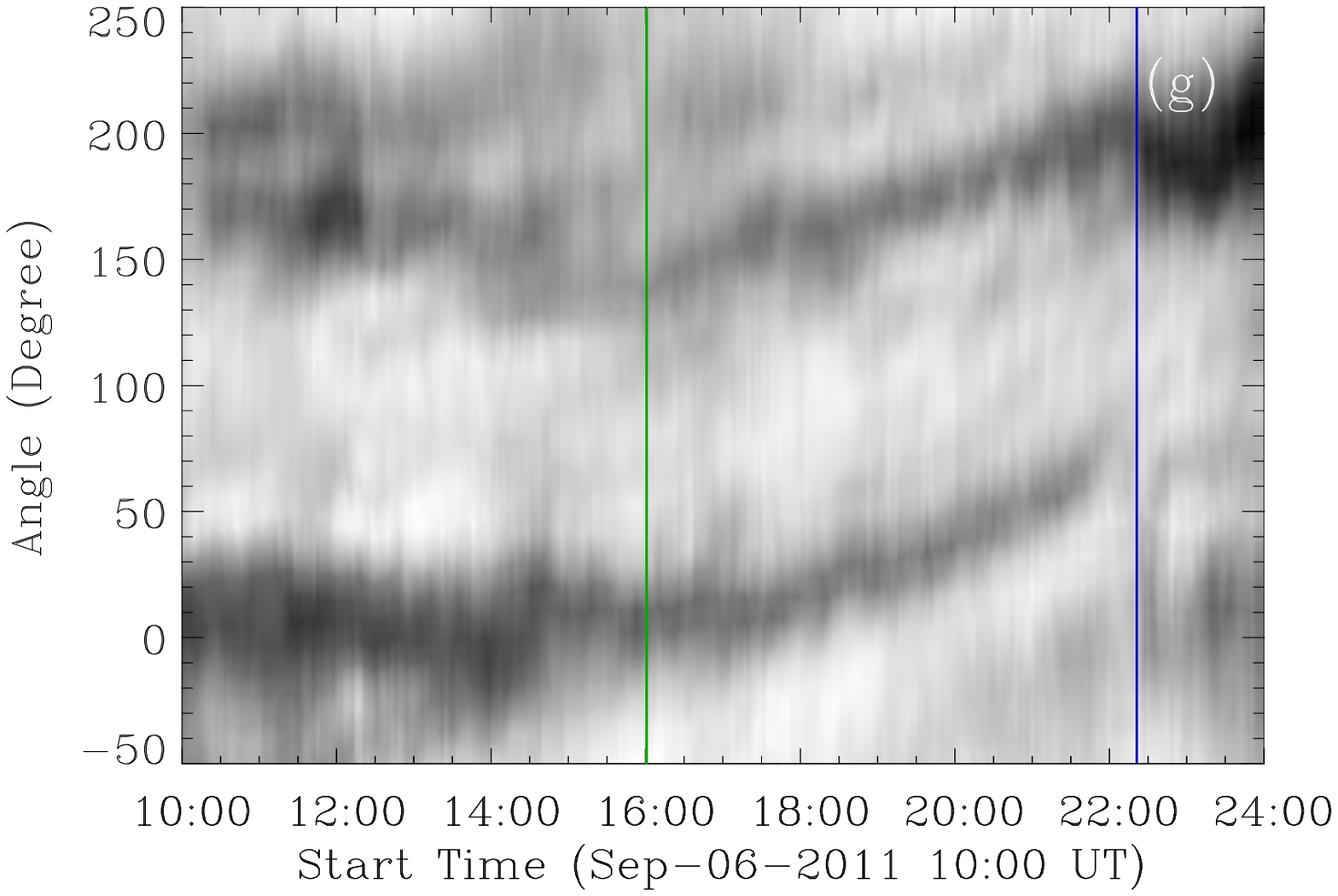}
\hspace{-6.0mm}
\includegraphics[width=0.42\textwidth,angle=0,clip]{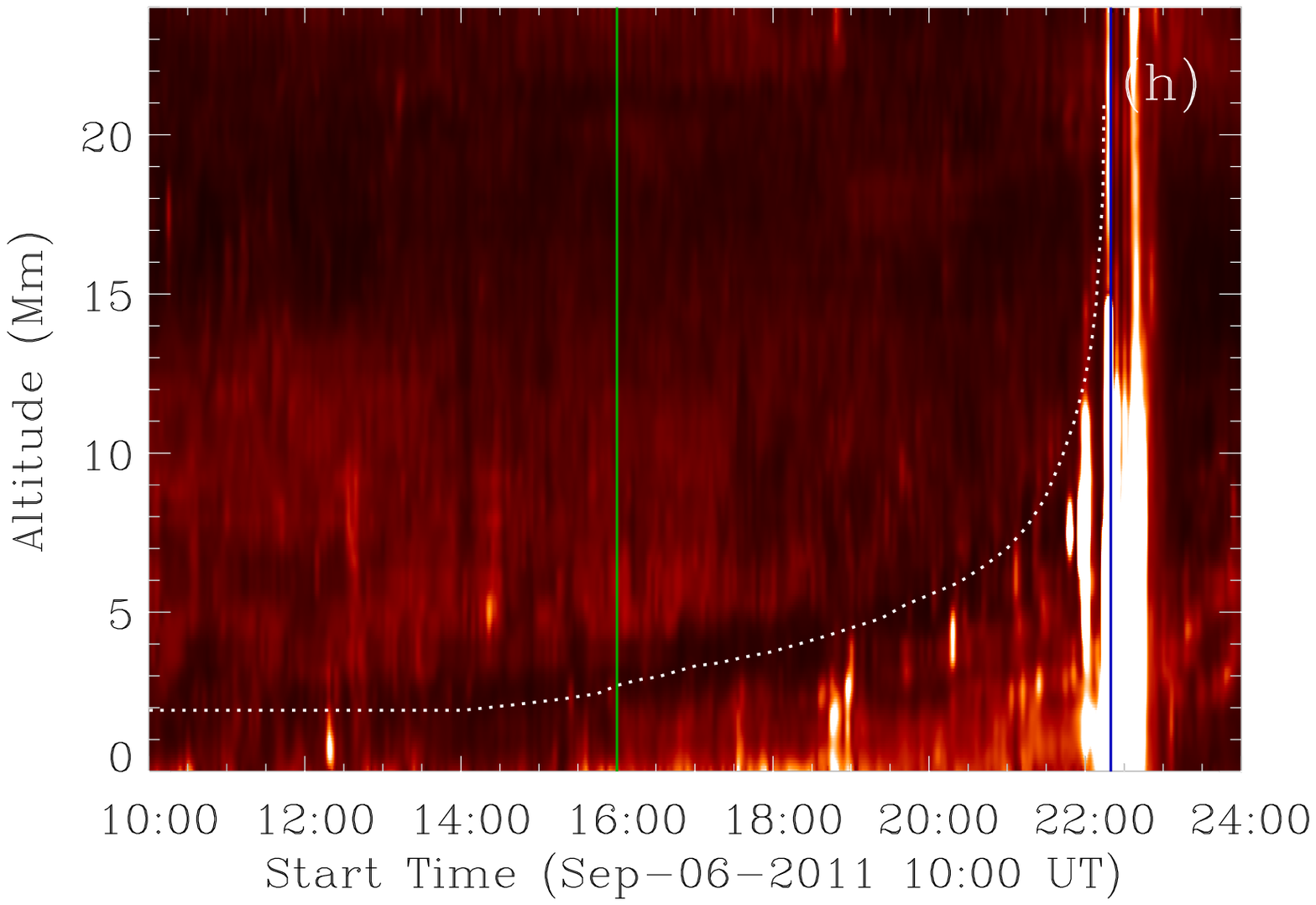}
\end{minipage}
\caption{(a-h): Sequences of sunspot (a-c) and filament (d-f)
morphological evolutions in the same FOV as that of Figure 1a. The
blue and green contours in panels b and e represent the $\pm$350~G
level of $B_z$ at $\sim$19:00 UT. The two arrows in panel a point
alongside the two magnetic tongues. (g): The $r-\theta$ plot of
the rotating sunspot. (h): The height-time plot for the filament
along the white line shown in panel d. The green and blue vertical
lines in panels g and h represent the start time of apparent
sunspot rotation (16:00 UT) and the flare peaking time (22:20 UT).
An animation and a color version of this figure are available
online.} \label{}
\end{figure*}

\begin{figure*}[!htbp]
\centering
\includegraphics[bb=40 140 450 550, width=80mm]{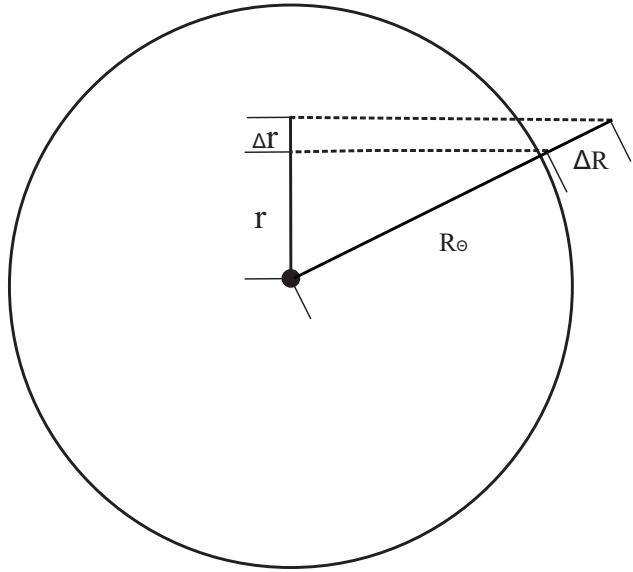}
\caption{ Schematic showing the relationship between the projected
and deprojected (i.e., real) filament heights ($r$ and $R$) and
rising distances ($\vartriangle$$r$ and $\vartriangle$$R$). See
text for more details. } \label{fig:bright}
\end{figure*}

\begin{figure*}[!htbp]
\begin{minipage}{\textwidth}
\centering
\includegraphics[width=170mm]{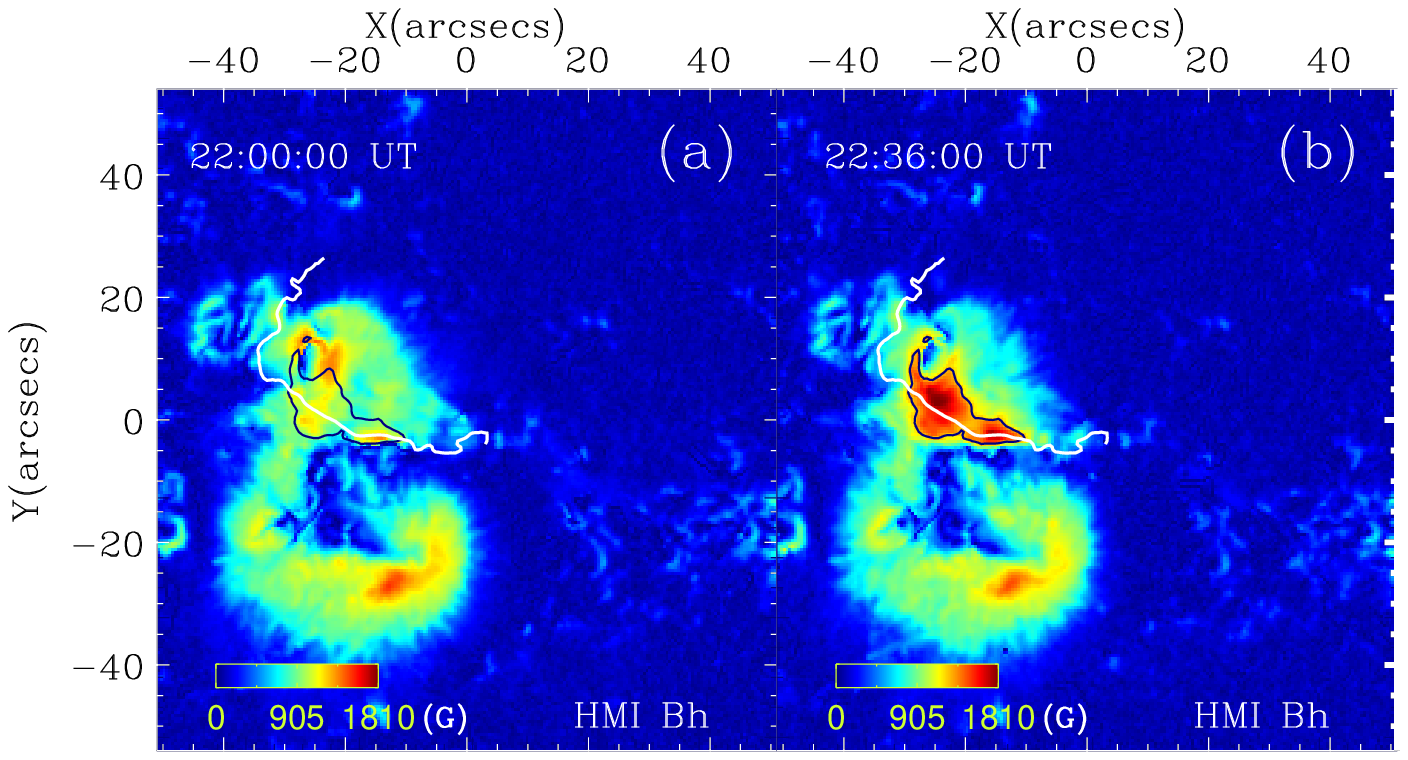}
\vspace{-12.3mm}
\end{minipage}
\hspace*{12.2mm}
\begin{minipage}{0.5\textwidth}
\centering
\includegraphics[width=71.55mm,clip]{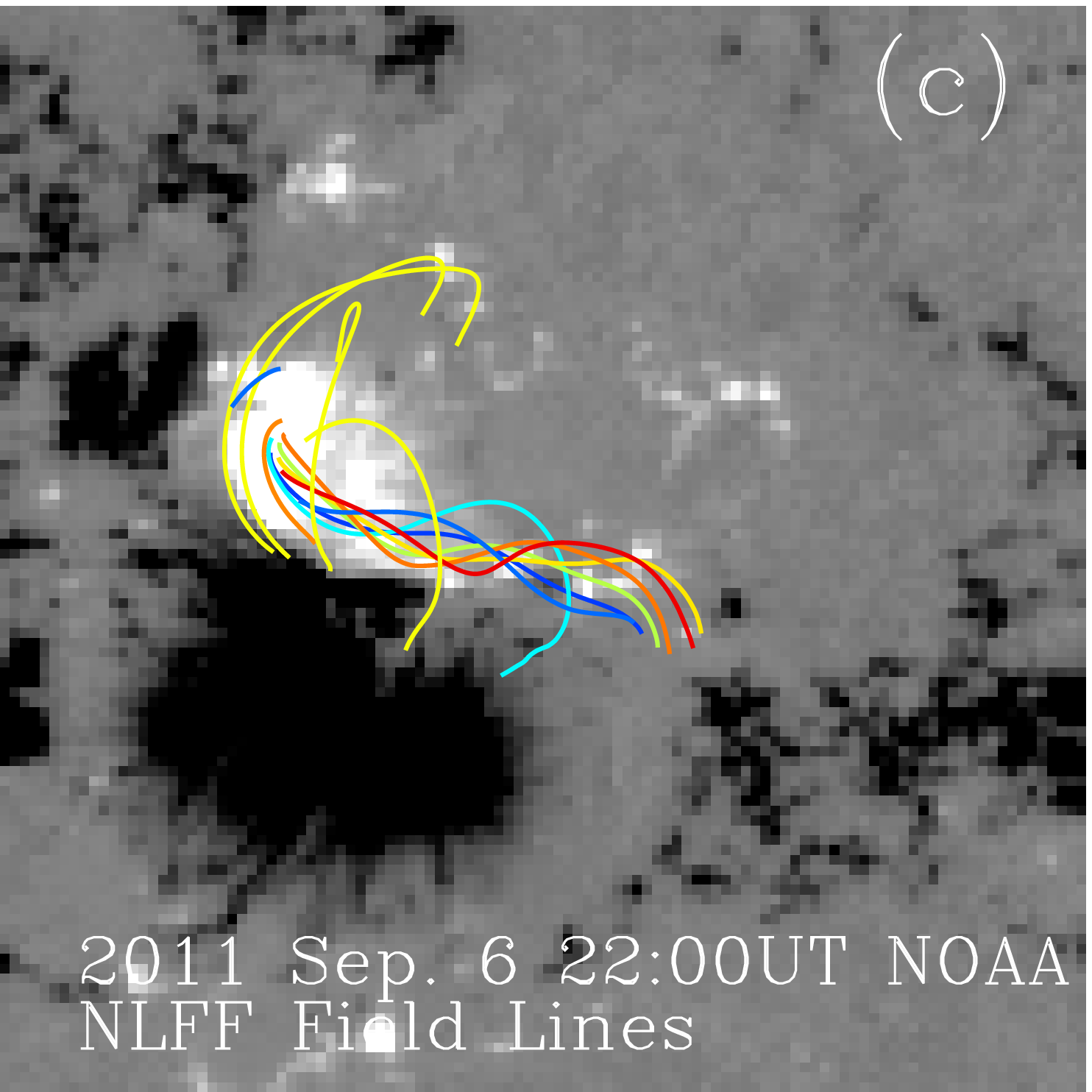}
\end{minipage}
\hspace{-12.8mm}
\begin{minipage}{0.5\textwidth}
\centering
\includegraphics[width=71.55mm,clip]{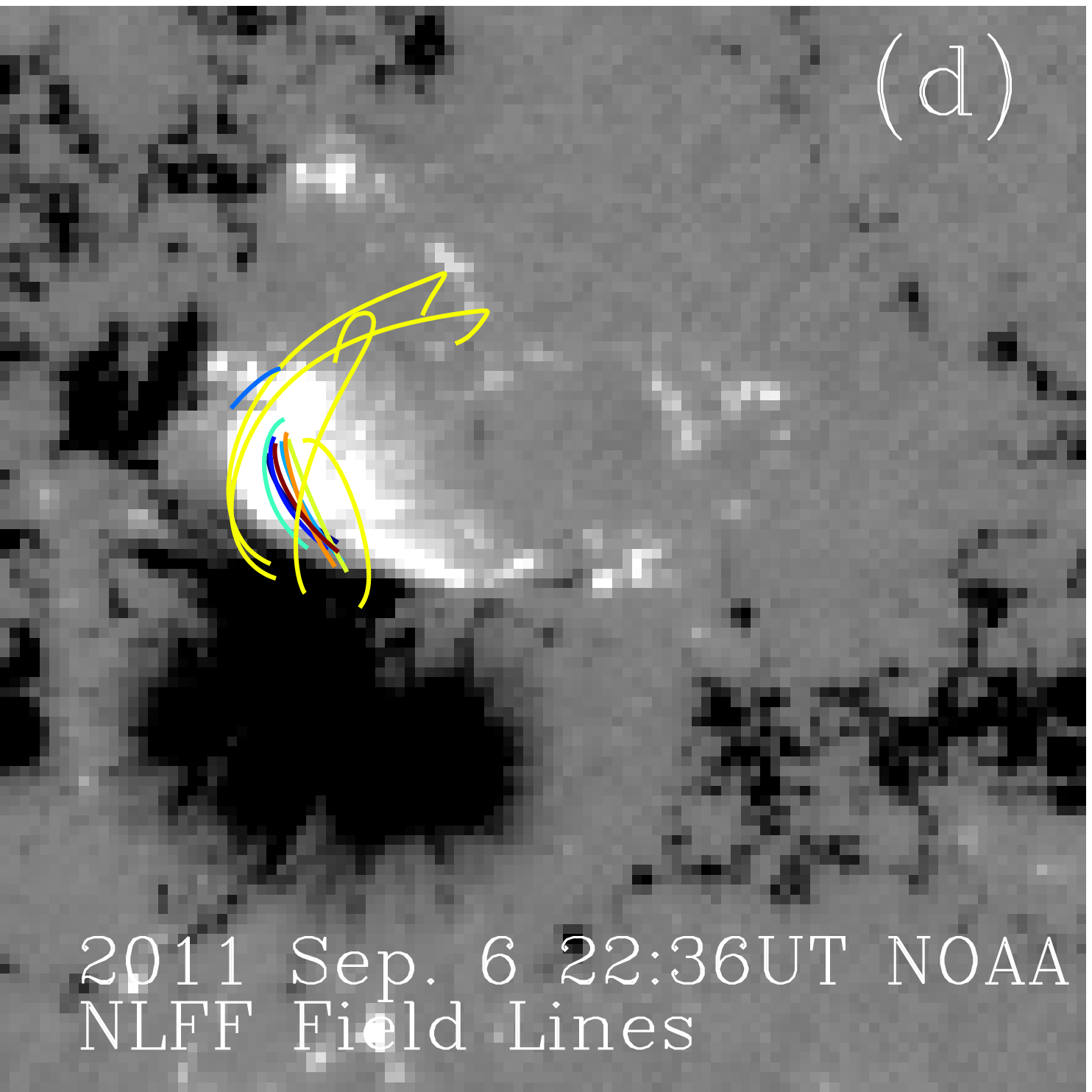}
\end{minipage}
\caption{(a, b): The $B_h$ distribution at 22:00 UT and 22:36 UT.
(c, d): Selected coronal field lines given by the NLFFF
reconstruction method. A color version of this figure is available
  online.} \label{}
\end{figure*}

\begin{figure*}[!htbp]
\centering
\includegraphics[width=150mm]{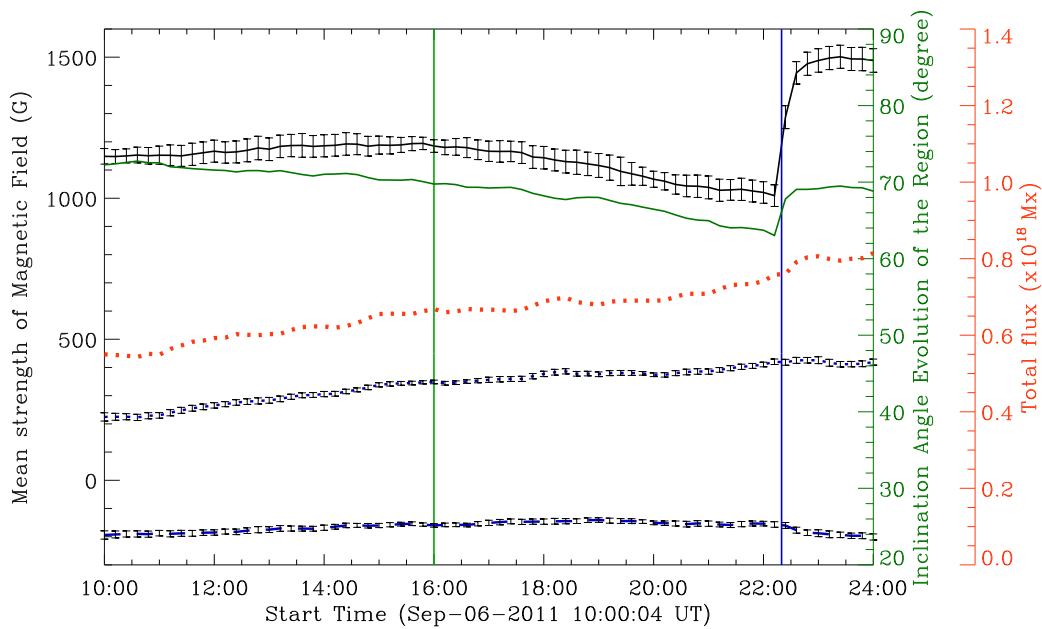}
\caption{The temporal profiles of the average $B_h$ (black solid),
the average of the positive (blue dotted) and negative (blue
dashed) $B_z$ components, the total flux (red dotted) and
inclination angle $\theta_B$ (green solid) in the area defined by
the black contour of Figure 5b. The green and blue vertical lines
represent the start time of apparent sunspot rotation (16:00 UT)
and the flare peaking time (22:20 UT). A color version of this
figure is available online.} \label{fig:bright}
\end{figure*}

\begin{figure*}[!htbp]
\hspace*{2.2mm}
\begin{minipage}{\textwidth}
\centering
\includegraphics[width=90mm, height=63mm]{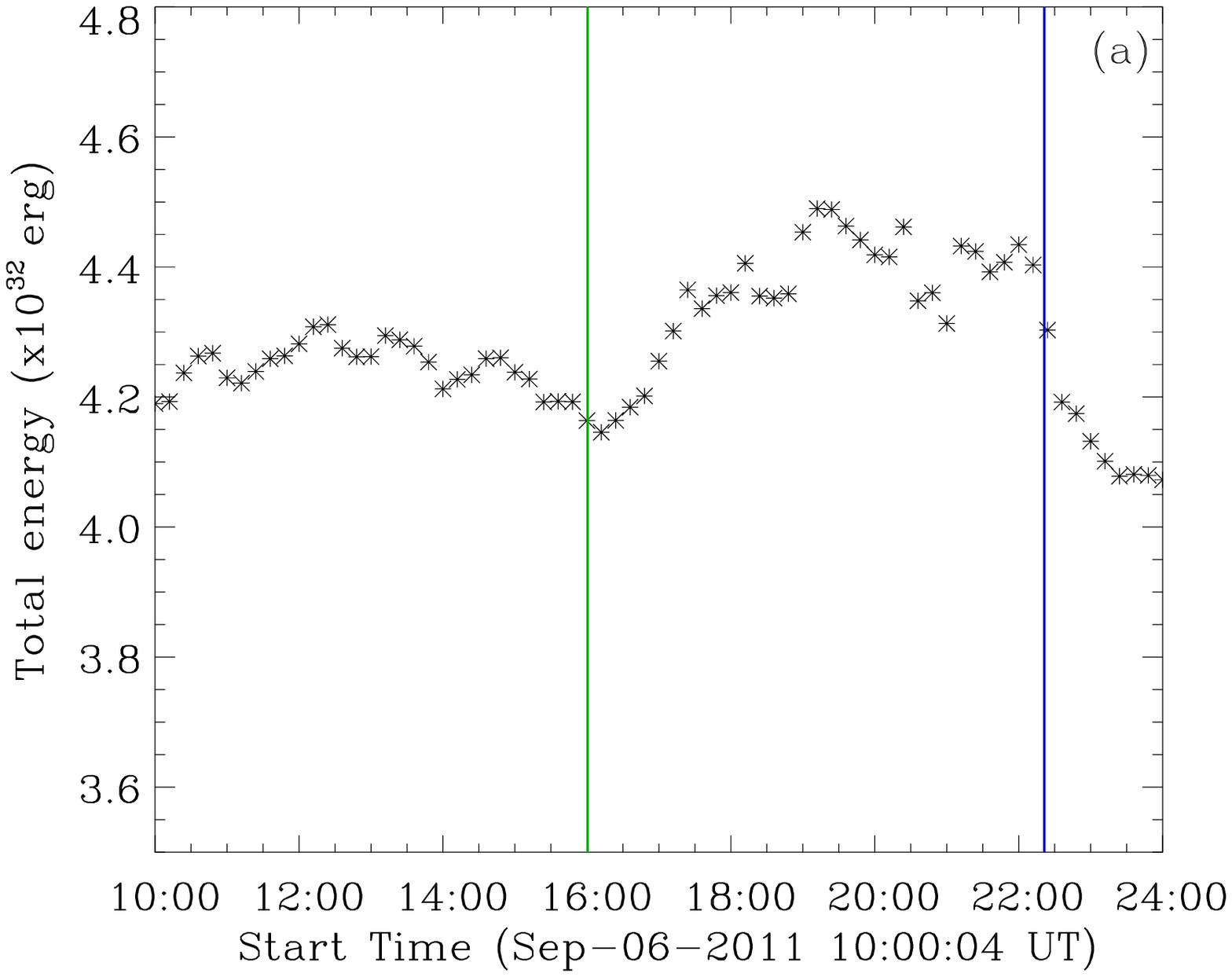}
\vspace{2.3mm}
\end{minipage}


\hspace*{1.0mm}
\begin{minipage}{\textwidth}
\centering
\includegraphics[width=90mm, height=63mm]{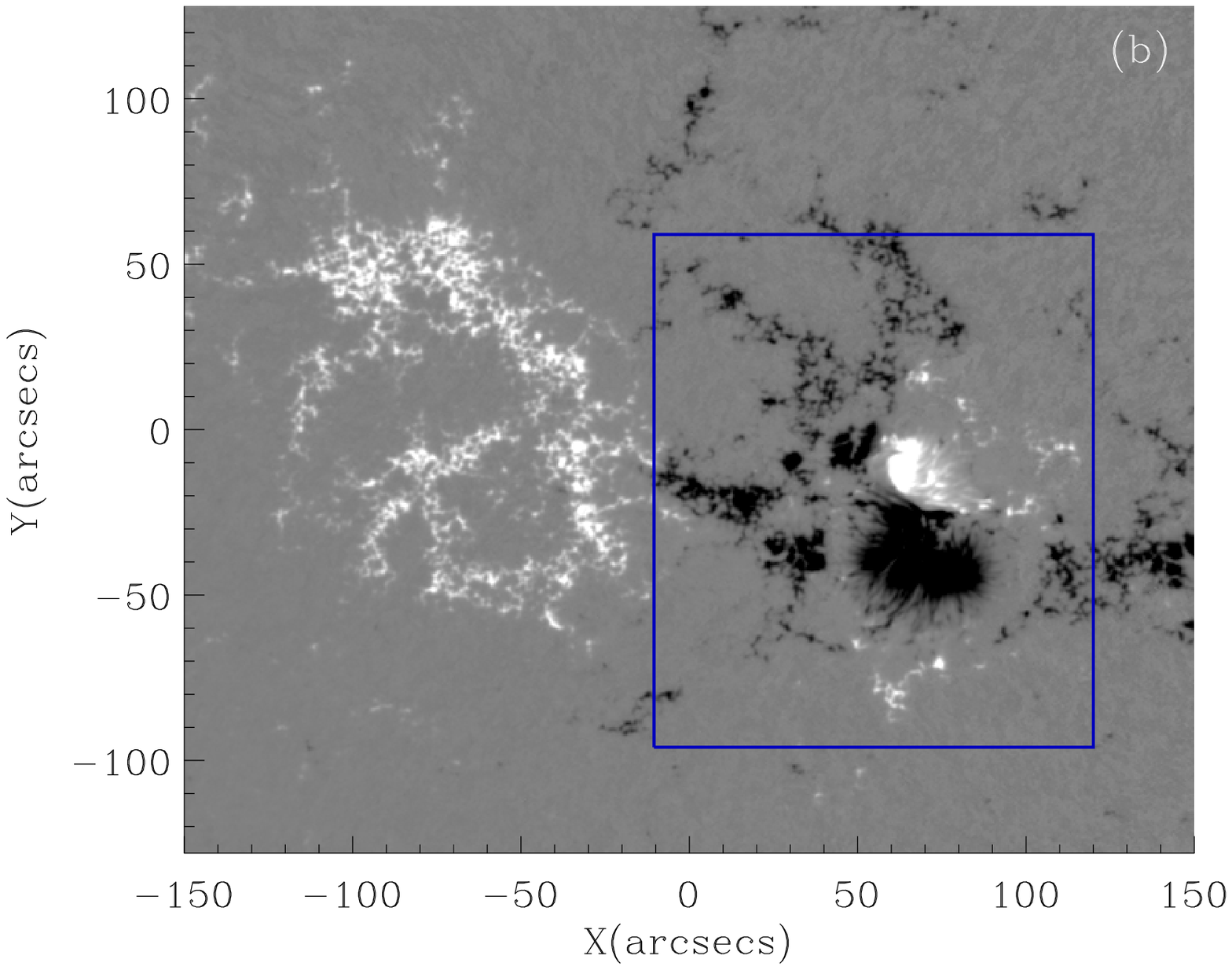}
\end{minipage}
\caption{The temporal profile (panel a) of the total energy of the
reconstructed magnetic field in a sub-volume with a bottom shown
as the blue square in panel (b) and the same height as that used
for the NLFFF reconstruction. The green and blue vertical lines
represent the start time of apparent sunspot rotation (16:00 UT)
and the flare peaking time (22:20 UT). A color version of this
figure is available online.} \label{fig:bright}
\end{figure*}

\begin{figure*}[!htbp]
\begin{minipage}{\textwidth}
\centering
\includegraphics[width=100mm, height=60mm]{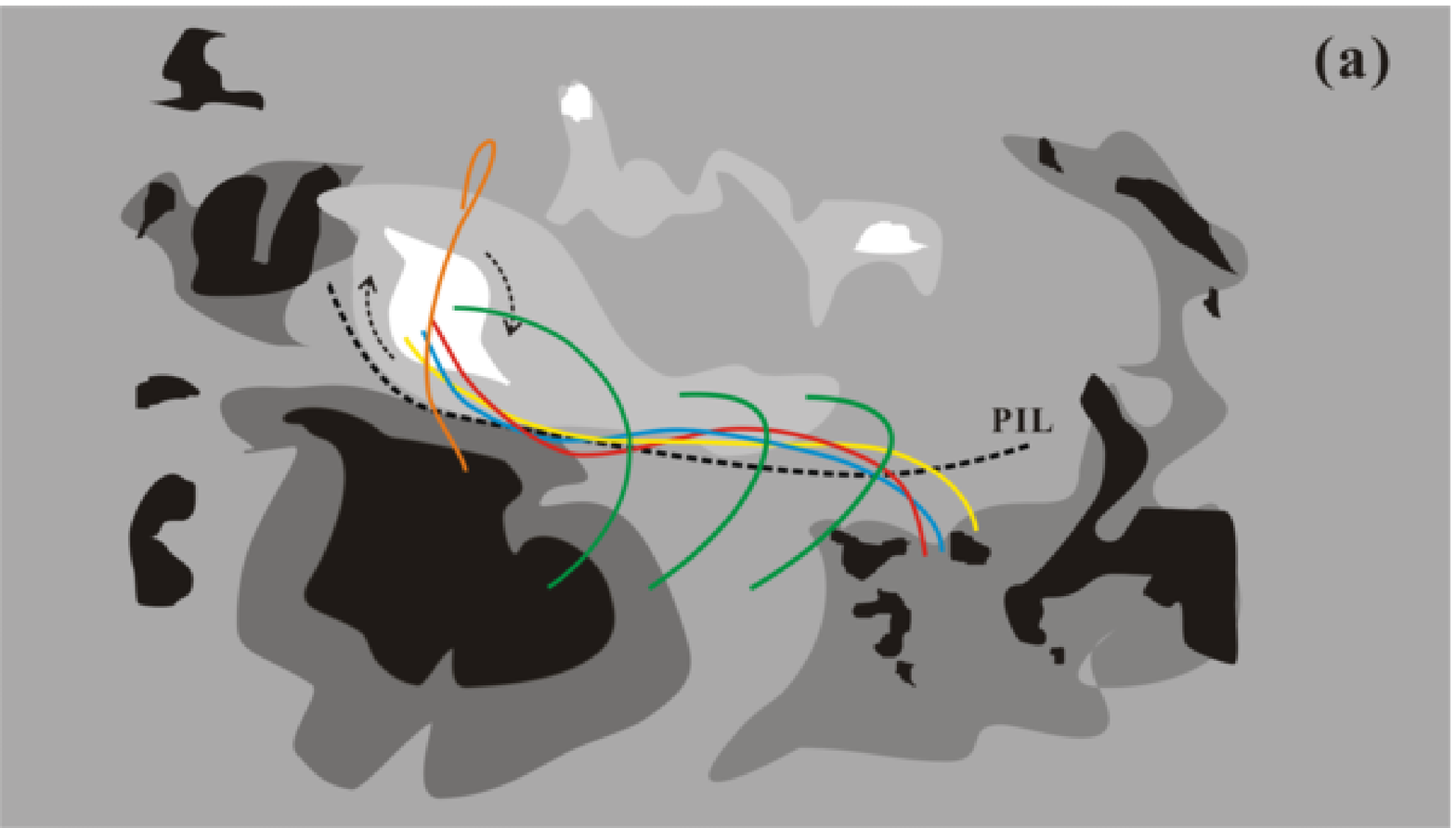}
\vspace{2.3mm}
\end{minipage}
\hspace*{4.4mm}

\begin{minipage}{\textwidth}
\centering
\includegraphics[width=100mm, height=60mm]{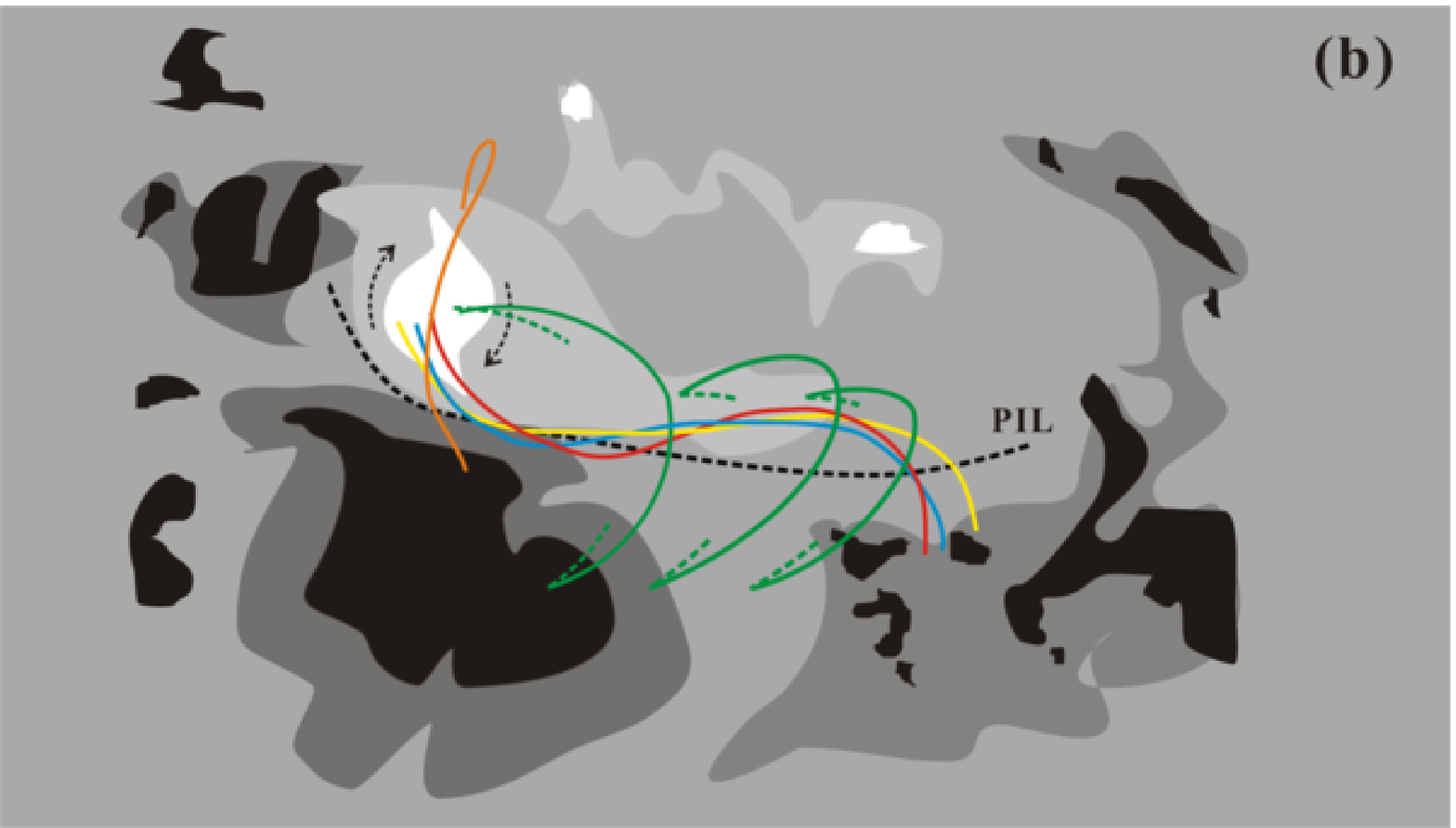}
\vspace{2.3mm}
\end{minipage}
\hspace*{4.4mm}

\begin{minipage}{\textwidth}
\centering
\includegraphics[width=100mm, height=60mm]{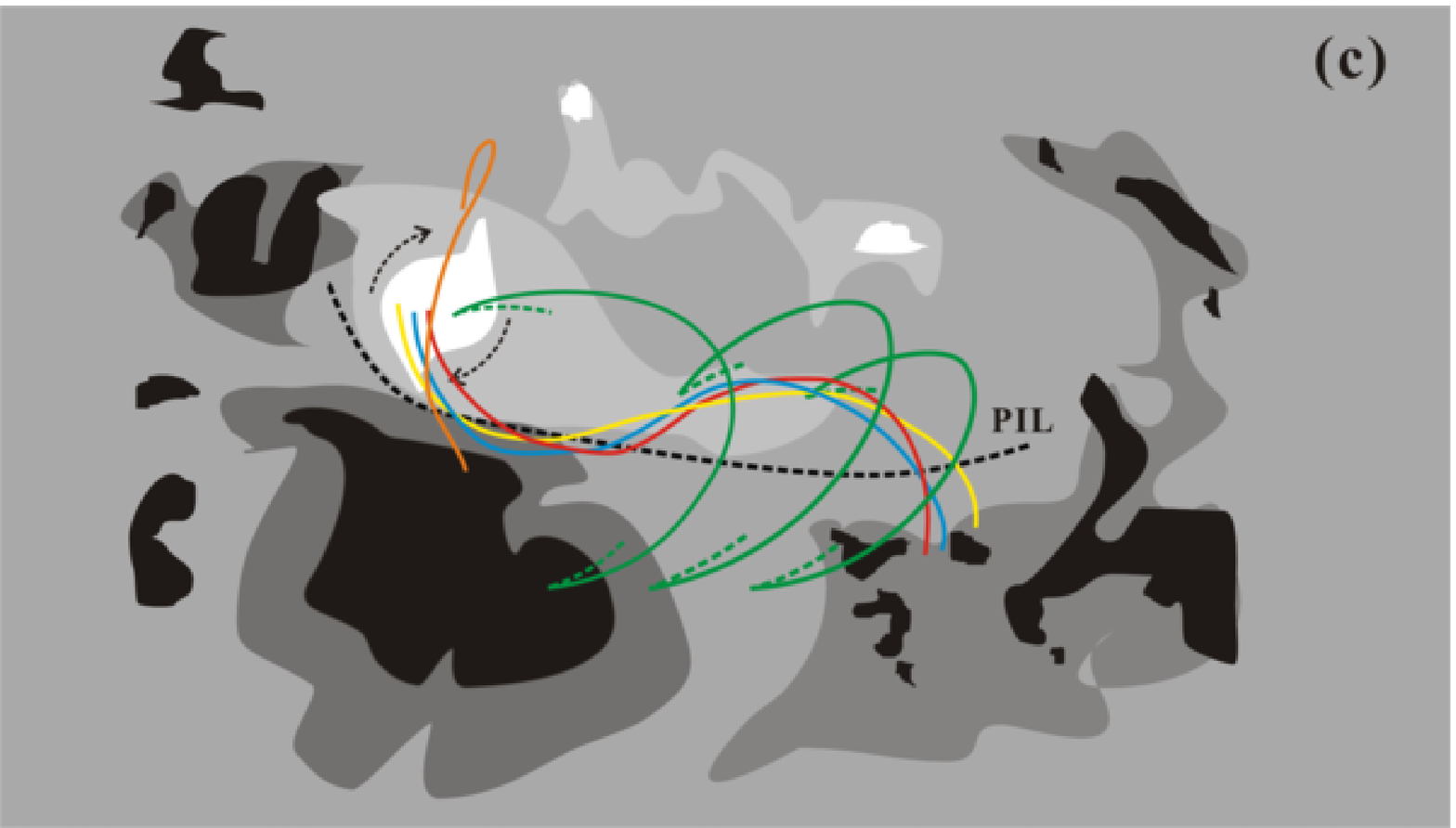}
\end{minipage}
\caption{Schematics of a flux rope CME driven by persistent
sunspot rotation. The rotating sunspot is indicated by the white
structure with two extending tongue structures. The rotating
direction is denoted by two curved arrows. The short green dashed
lines indicate the field line location at the preceding moment.
See text for more details. A color version of this figure is
available online.} \label{}
\end{figure*}

\end{document}